\documentstyle[12pt,aps,epsf]{revtex}

\input{epsf}
\begin{document}

\author{Kozlov G.G. {\it e-mail:} \hskip 4pt gkozlov@photonics.phys.spbu.ru}

\title{Simple device for scanning image}
\maketitle

\vskip30pt

\begin{abstract}
The simple device for scanning image is described. It
has much in common with usual TV camera, with
an electron beam replaced by an optical one. After the general description
of the device, we present a simple experimental illustration.
\end{abstract}

\section{General description}

In this letter we suggest a simple device for scanning image, which has
much in common with usual TV camera with an electron beam replaced by an
optical one. The scheme of the device is presented in Fig.~1, where (1) is
the transparent electrode; (2) is the semiconductor layer with the gap, {\it
wide enough to be transparent} for the light from the image; (3) is the
semiconductor layer with the gap, {\it narrow enough  to be not
transparent} for the light from the image, and (4) is the electrode, or
the substrate with high conductivity. Suppose, that some image is formed
in the plane of the layer (3) by means of some optical system. Though the
image produces the profile of photoconductivity in the layer (3), there is
no current in the circuit, because the layer (2) is still an isolator ---
note, that the layer (2) is transparent for the image light, which causes
no photoconductivity in this layer. Now, let us switch on the narrow {\it
reading beam} of a short enough wavelength, to produce the
photoconductivity in the layer (2). This beam produces the narrow path (5)
with high conductivity in the layer (2), which gives rise to the current in
the circuit. The value of this current is proportional to the local
illumination intensity of the image. For the sake of simplicity we suppose,
that the reading beam has absorbing length much shorter, than the thickness
of the layer (3), so it does not produce any noticeable photoconductivity
in the layer (3). Now, if we move the reading beam along the image, the
current in the circuit will drop down, when the reading beam is passing
through the dark areas of the image, and will increase, when the beam is
passing through the bright areas. This process is similar to that in
conventional TV camera, with the optical beam instead of the electron one.

\section{Experimental illustration}

To test the aplicability of the above scheme, we construct the simple
installation shown in Fig.~2. The exact scheme of the above device is
presented in the inset.  Transparent electrode (1) and electrode (4)
(notations of Fig.~1) are the water layers.  The 3 mkm AlGaAs layer is
playing the role of the layer (2) (see Fig.~1) with the relatively wide gap
($\sim$1.9 eV). And finally, the 300 mkm of semiconducting GaAs substrate
is playing the role of the layer (3) with the narrow gap ($\sim$1.5 eV).
The beam of the HeNe laser (0.63 mkm, or 1.97 eV), which plays the role of
a reading beam, passes through a simple deflecting system (consisting  of
the loudspeaker, the mirror (1) and the lens (see Fig.~2)), and gives a
small spot moving along the linear segment in the plane of the device.

The Nd laser beam (1.06 mkm, or 1.17 eV) formes a small spot in the same
plane, which plays the role of the image. Manipulating with the mirror (2)
one can force the reading beam to pass through this spot. When this takes
place, the short pulses of current occur in the circuit. It should be
pointed out that despite GaAs is transparent for the 1.06 mkm radiation,
the photoconductivity still occur, due to the presence of the impurities
and intrinsic defects in the semiconducting GaAs substrate.

The main property of the described device is its simplicity. This device
may be preferable in all cases, when high energy electron beam can damage
the light-sensitive surface. Also, the device of this kind can perform some
integral transformations, similar to the Fourier transformation. For this
purpose one should substitute the narrow running reading beam with the
sinusoidal profile, which gives the sinusoidal distribution of the
conductivity in the layer (2).

\vskip 5mm
The author is not sure, that such a simple device was not described, or
used before, and will be gratefull for any references and coments.

\begin{figure}
\epsfxsize=600pt
\epsffile {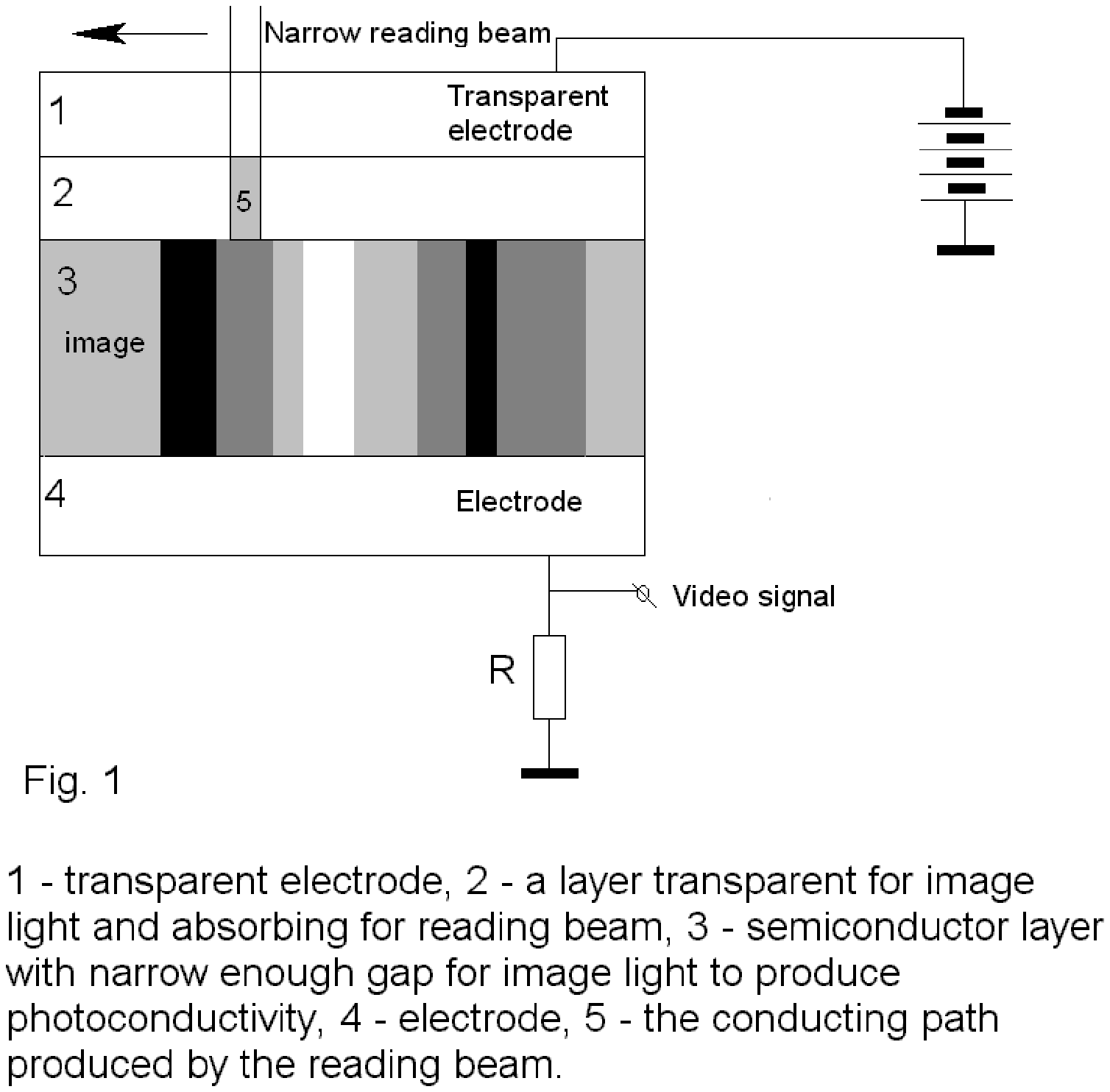}

\end{figure}

\begin{figure}
\epsfxsize=600pt
\epsffile {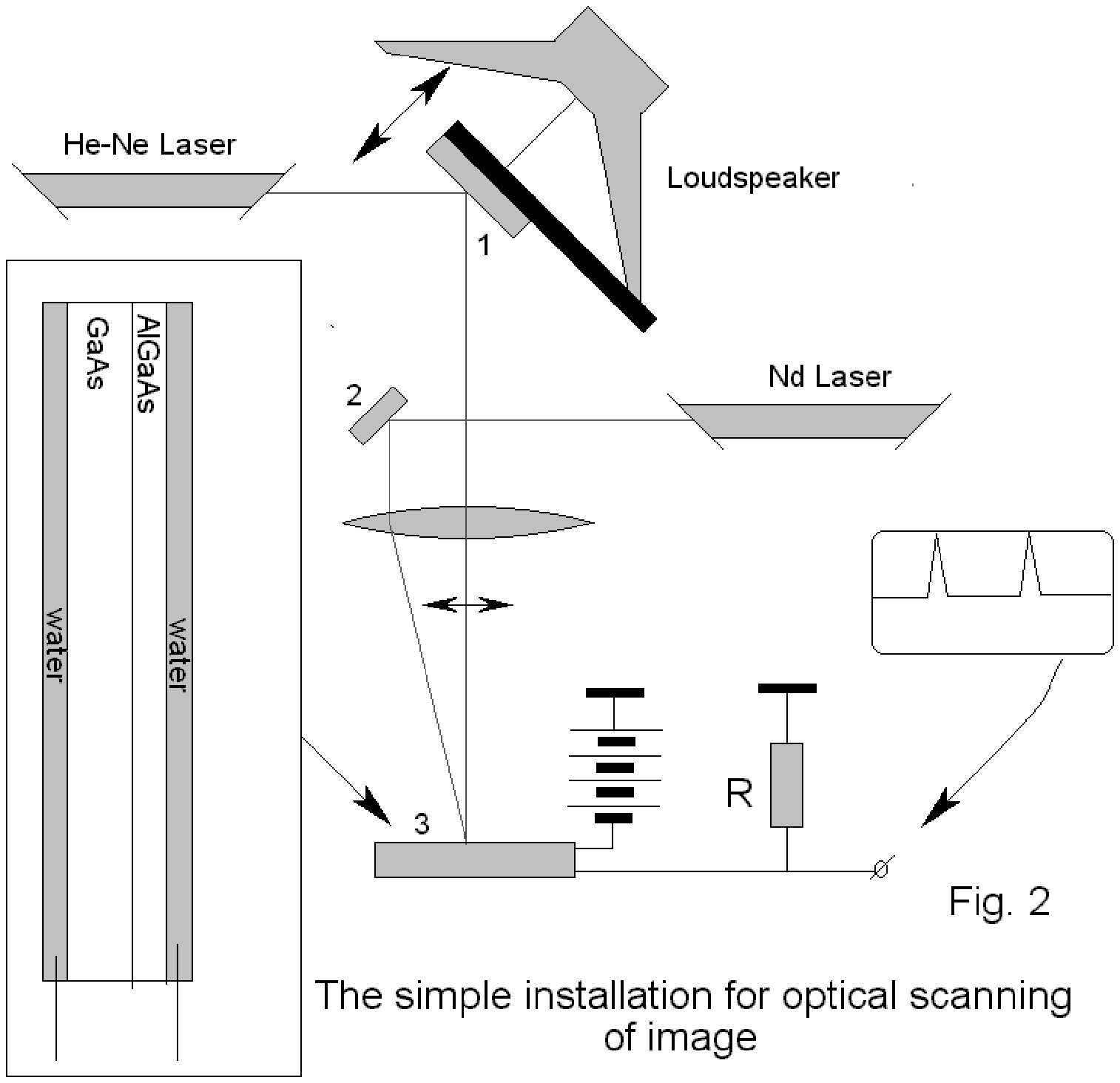}

\end{figure}

 \end{document}